\documentclass[preprint2]{aastex}

\shorttitle{Multiple  supra--arcade reconnection outflows in solar flares}
\shortauthors{C\'ecere et al.}
\usepackage{txfonts}

\begin{document}
\title{Simulation of descending multiple  supra--arcade \\ reconnection outflows in solar flares}
\author{M. C\'ecere\altaffilmark{1,2}, M. Schneiter\altaffilmark{1,3,4}, A. Costa\altaffilmark{1,3,4}, S. Elaskar\altaffilmark{1,4} and S.  Maglione\altaffilmark{5}}
\altaffiltext{1}
{Consejo Nacional de Investigaciones Cient\'\i ficas y T\'ecnicas (CONICET),
  Argentina.}
\altaffiltext{2}{Instituto de F\'\i sica Enrique Gaviola, IFEG--CONICET,  C\'ordoba,  Argentina} 
\altaffiltext{3}{Instituto de Investigaciones en Astronom\'\i a Te\'orica y Experimental
IATE--CONICET,  C\'ordoba, Argentina. }
\altaffiltext{4}{Facultad de Ciencias Exactas, F\'\i sicas y Naturales, Universidad Nacional de C\'ordoba (UNC),  C\'ordoba,  Argentina}
\altaffiltext{5}{Facultad de Ingenier\'\i a, Universidad Nacional de R\'\i o Cuarto, Ciudad Universitaria, R\'\i o Cuarto, C\'ordoba, Argentina}

\begin{abstract}
After recent AIA observations by Savage, McKenzie and Reeves we revisit the scenario proposed by us in previous papers. We have shown that sunward, generally dark plasma features
  originated above posteruption flare arcades 
are consistent with a scenario where plasma voids (which we identify as supra--arcade reconnection outflows, SAROs) generate  the bouncing and interfering of shocks and expansion waves upstream of an initial localized deposition of energy which is collimated in the magnetic field direction. In this paper we analyze the multiple production and interaction of
SAROs and their  individual structure that make them relatively stable features while  moving. We compare our results with  observations and with the scenarios proposed by other authors.

\end{abstract}

\section{Introduction}\label{Intro}
Sunward dark moving trails with origin [$ 40-60$]Mm
  above posteruption flare arcades and decelerating speed in the
  range $\sim [50 - 500] $km s$^{-1}$ were first detected with the
  \textit{Yohkoh} Soft X--ray Telescope (SXT).  Since then, they
have been extensively reported using other
instruments such as \textit{TRACE} \citep{2003SoPh..217..267I, 2003SoPh..217..247I},
\textit{SOHO}/SUMER \citep{2003SoPh..217..247I} and \textit{SDO}/AIA \citep{2012ApJ...747L..40S}, \citep{2011ApJ...742...92W}.
The lack of X--ray and extreme--ultraviolet (EUV)
   signatures in images and spectra has lead to consensus  that
   these downward moving structures are likely  voided
   flows generated by reconnection processes in a current sheet above
   the flare arcade.  Besides the dark moving structures, bright
   supra--arcade downflowing features have also been reported during
   flares \citep{2000SoPh..195..381M}. 

\citet{2009ApJ...697.1569M}, \citet{2011ApJ...730...98S} suggested that
  supra--arcade downflows (SADs) are the cross--sections of thin and
  empty flux tubes retracting from a reconnection site high in the
  corona. According to these authors, the high enough inner magnetic
  pressure could be the reason  the voids are able to resist being filled
  in immediately by the surrounding denser plasma. 
 \citet{2009EP&S...61..573L} proposed a scenario where the dynamic of retracting
 magnetic fields is triggered by a localized reconnection event that
 produces up and down flowing reconnected flux tubes,
which are slowed down by underlying magnetic arcade
   loops. The observed SAD speeds are lower than
 expected for reconnection outflows in regions of typically assumed Alfv\'en speeds
of $1000$ km s$^{-1}$. \citet{2009EP&S...61..573L} suggested that drag forces
 could work against the reconnection outflow.

  \citet{2005A&A...430L..65V} analyzed TRACE downflow oscillations transverse to the
  magnetic field.  They found that the initial speeds and the
  displacement amplitudes, of  kink--like type in the observational
  dark lanes of variable sizes (between [$ \sim 2-9$]Mm), decrease as
  they propagate downwards, while the period remains fairly constant with
  height.

  Recently, after AIA observations with high resolution and cadence,
  \citet{2012ApJ...747L..40S} re--interpreted SADs as wakes
  created by the retraction of thin loops (SADLs) instead of
    the previous interpretation as flux tube cross--sections.   SADLs are features of sizes [$
    \sim 0.9-1.3$]Mm observed during the early phase of the eruptive
  event and SADs as features of sizes $ \sim 9$Mm that become
  apparent afterwards, in contrast to bright, high--temperature plasma associated with current sheets. They proposed that deceleration is expected due
  to the buildup of downstream magnetic pressure and/or drag
  mechanisms.

In \citet{2009MNRAS.400L..85C}, \citet{2010MNRAS.407L..89S} and \citet{2011A&A...527L...5M} (hereafter   Paper 1--3, respectively) we showed, by means of 1 and 2D MHD
  simulations, that the dark tracks are consistent with plasma voids
 generated by the bouncing and interfering of shocks and expansion
  waves upstream of the initial localized deposition of energy.
 The
composition of both, a resulting sunward directed hydrodynamic shock
pattern and a perpendicular magnetic shock, produce an overall
transversely shaking void that propagates towards the surface of the sun,
 resembling the kink--like mode described in \citet{2005A&A...430L..65V}. Contrary to the 1D results --where the sunwards dynamic is  
independent of the magnetic field intensity owing to its exclusive
waveguide role-- in the 2D simulation the sunwards speed is higher
for higher values of the magnetic field.  This can be interpreted as the capability of the
low coronal plasma to collimate the deposition of energy in the
magnetic field direction. 

 As the obtained  $\beta$ values were larger than those outside  the
 voided cavity (see Fig. 2 of Paper 2), with the contour in total pressure equilibrium, we
 concluded that the internal magnetic pressure cannot be responsible
 of preventing the collapse of the vacuum
 zone. 
Instead, we found that the higher values of $\beta$
  inside the dark lanes are due to  hotter plasma --with  low inner values ​​of
    density ($\sim 2 \ 10^{7}$ cm$^{-3}$)-- concentrated in regions 
  of $T > 20$ MK \citep{1999ApJ...527L.121W}. 
Actual  observations do not show significant emission values of the voided plasma in any of the instrumental wavelengths (e.g., high--temperature AIA bandpasses). Hence, the observations could  be consistent   with these plasma parameters.

According to  \citet{2012ApJ...747L..40S}  shrinking loops or SADLs are always leading SAD features.  Both SADs and SADLs are interpreted as the outflows created during the re--organization of the magnetic field during the reconnection process.
However, descending SADs are only seen trailing  loops once the current sheet has become noticeably filled with plasma.  In this paper, in  light of the new AIA observations we revisit the scenario proposed in Papers 1--3. 
We  simply reinterpret our previous 1D and 2D resulting voided features as supra--arcade reconnection outflows (SAROs)   to differentiate them from the SAD description  as stated by  \citet{2012ApJ...747L..40S}. In our 2D model we are not simulating  the arcade loops nor the possible interaction of the SAROs  with the loops, and  will come back to the SAD and SADL features later.  Hence, reconnection is modelled as a pressure pulse representing  the energy deposition that
  triggers the phenomenon that creates the dark track which we call SAROs.

  With the aim of testing the scopes of our model, we first focus on
  reproducing the dynamic and oscillating interaction between several of these SARO voided
  structures. 
In Paper 1--3 we found that to reproduce the oscillating pattern we required closed boundary conditions. However, in Paper 2 we found that the oscillating pattern was robust under a broad  variation of  the boundary conditions --contemplating different density values at the extremes and allowing total or partial boundary rebounds.  Thus, the question arises about  the role played by the boundary conditions.
The observations show that the outflows sustain their individual configuration all the way down into the arcade structure. Thus, the simulation  of multiple outflows   triggered by different pressure pulses  must maintain these individualities while they  evolve decelerating sunwardly. However, can we consider the dynamic of a multiple outflow structure  as the superposition of individual ones (as described in Paper 1--3)?
If this is the case,  what produces the wavy features described in \citet{2005A&A...430L..65V}?  

  \section{Numerical code and initial conditions}  
We carried out the 2D-MHD simulations with the Mezcal
  code, which has been extensively tested in several scenarios (see
  for instance \citet{2006A&A...449.1061D} or \citet{2008ApJ...689..302D}).  
All calculations were
performed with a numerical grid of $(x,y)=(400,800)$ grid--points
and a physical size of $(20,40)$Mm, corresponding to a resolution of $(50,50)$km. The coordinate $y$
represents the sunwards direction and the $x$ coordinate the
transverse to the magnetic field one. We
assumed a constant radial initial magnetic field structure, and typical
background temperature of $T=1.0 \ 10^{6}$K. 
Simultaneous ($t=0$sec) spherical pulses
$(\Delta P/P)_{i}=(110,90,110) \ i=1,3$ ($ \Delta P$ is the triggering pressure pulse and $P$ is
the background gas pressure of the corona) of a radius $0.6$Mm were
localized in positions $(67,720),$ $(167,680)$ and $(333,720)$
corresponding to $(3.35,36.0)$Mm, $(8.35,34.0)$Mm and $(16.65,36.0)$Mm,
respectively.

 Several simulations were carried out, varying the magnetic 
 field, $B $ and the density $\rho$. Also, we added a localized deposition of energy -modeled as a new  triggering pressure pulse, $(\Delta P/P)_{4}$- resembling  a new reconnection event (at $t=200$sec) occurring in the scenario modified by the later ones. 
Table \ref{tab:table1} shows  the  models that we use to discuss our results. 
We employed  open and closed  conditions 
 for the lateral boundaries, whereas, initially, 
  only a fix rebound condition was used at the upper radial direction, to simulate the outflows 
  resembling the action of the reconnection site, and transmission
  condition for the bottom boundary, assuming that the perturbations
  are absorbed in the sunwards direction. Finally, we also used  open boundary conditions for the upper radial direction with a non--homogeneous initial background density ($n = n_{0}   \sin(\pi x / 2) + n_{1}$, $(n_{0},n_{1}) =(2.07, 2.53) \ 10^{9}cm^{-3}$, $x$ in Mm) to emulates the passage of previous SARO structures  and to explore other plasma configurations. The characteristic boundary
  parameters were chosen in accordance with typical observed dark lane
  structures, as in \citet{2005A&A...430L..65V} and \citet{1999ApJ...519L..93M}.

\begin{table}
\begin{tabular}{cccc}
 \centering
Model&$B [G] $& $n*10^{9}[cm^{-3}]$& $(\Delta P/P)_{4}$ \\ \hline  
$M1$&$3$&$0.46$& $400$ \\ \hline
$M2$&$10$&$0.46$& $400$ \\ \hline
$M3$&$10$&$3$& $400$ \\ \hline
$M4$&$3$&$0.46$& $200$ \\ \hline
 \hline
\end{tabular}
\caption{\label{tab:table1} Simulated models:  $B$ the background magnetic field in the radial direction,  $\rho$ the background density.}
\end{table}%*}
\section{Results and Discussion}

\subsection{SAROs plasma  parameters}

\begin{figure*}%[htb!]
\centering
 \includegraphics[width=16.cm]{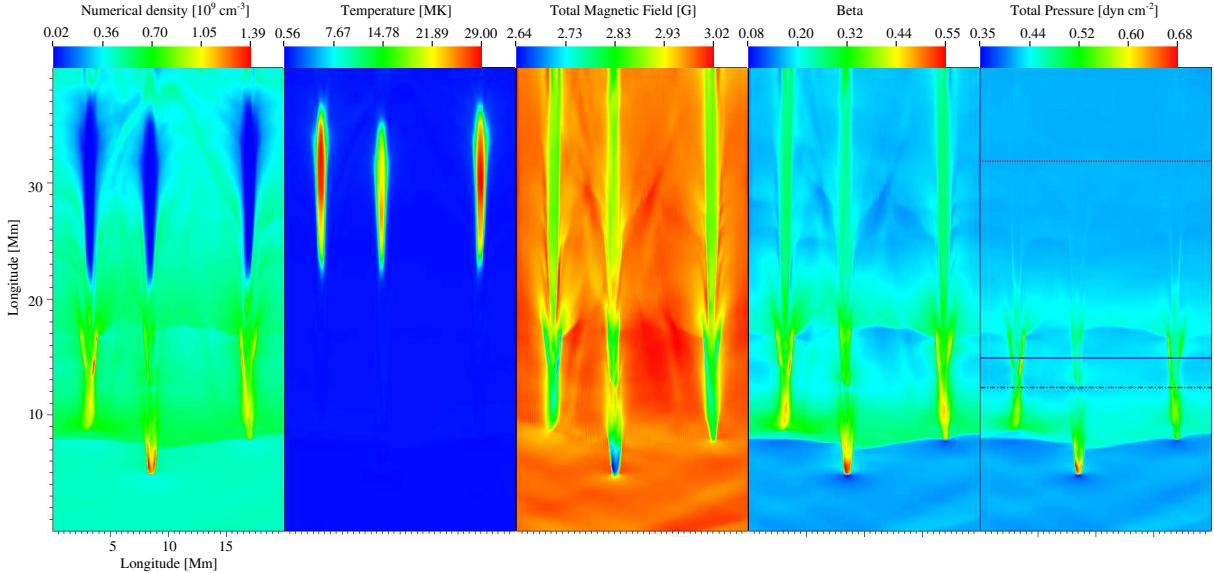}
 \caption{Simulation, at $t=196$sec, of: a) density, b) temperature, c) total magnetic field d) $\beta$--parameter and e) total pressure of the three outflows.}
\label{fig:uno} 
 \end{figure*}
Figure~\ref{fig:uno} shows, the $M1$ evolution of the three  voided outflows at $t=196$sec, it
simulates, respectively,  the voided density tracks left by  three simultaneous reconnection bursts,  the temperature, 
the magnetic field, 
the  $\beta$--parameter and the total pressure for open boundary conditions.
From  Fig.~\ref{fig:uno}a we see that, as expected (e.g. \citealp{1999ApJ...519L..93M})  the vacuum density values are less than the  environment ones
 in more than one order of magnitude (see the features above $y=20$Mm);  
the temperature  (Fig.~\ref{fig:uno}b) is more than one order of magnitude
higher in the voided cavities than outside (as in \citet{2009ApJ...690L..18L} and \citet{1999ApJ...527L.121W}); the inner magnetic field is lower than 
its  external value ($< 1$G) (Fig.~\ref{fig:uno}c),  thus,  the inner magnetic pressure cannot be responsible for preventing 
the collapse of the vacuum zone. 
The  $\beta$--parameter (Fig.~\ref{fig:uno}d) is larger inside than outside. 
Furthermore, from the upper part of Fig.~\ref{fig:uno}e we can see that  the total pressure of the region is almost constant, i.e.,  the features that characterize 
the vacuum outflows have disappeared,   meaning that the contour of the lanes are, on average, in total pressure equilibrium, as in Paper 3.

The Fig.~\ref{fig:uno} features  are  consistent with the dynamical  features described by \citet{2005A&A...430L..65V} except for the lack of oscillations in the 2D pattern and  the homogeneity of the background medium. 
The multiple outflow dynamics reproduces the individual dynamic; i.e., in accordance with observational data,   the temporal evolution analysis gave us 
decelerating  speeds of the order of $100-300$km s$^{-1}.$ The freezing--in of the plasma to the magnetic field  induces the collimation
of most of the kinetic energy towards the longitudinal direction, which  is enhanced with the increase of the magnetic field. In accordance with Paper 3 (see Table 1 of the paper)
--where we obtained   a slow increase in the initial  speed with the increase of
the background magnetic field-- for a run of $200$sec, varying the magnetic field from $3$G to $10$G (models $M1$ and $M2$) the distance evacuated by the perturbation increases  $4$Mm; while varying the background numerical density from $0.46 \ 10^{9}$cm$^{-3}$ to $3.0 \ 10^{9}$cm$^{-3}$  (models $M2$ and $M3$) the distance evacuated by the perturbation  diminishes in $6$Mm. 

At the initial stages  of the flare event, before the formation of fan rays, the supra--arcade zone can be assumed as an almost homogeneous medium.  Figure~\ref{fig:uno} intends to describe the initial dynamic of SAROs that descend from an apparent homogeneous background  medium towards the arcades.   Different   background density give dynamical  results consistent with the parameter range variations described by  \citet{2005A&A...430L..65V}. 
Thus, in the scenario proposed, the SARO  dynamics of several events is sustained by the interaction of localized (confined to the individual dark tracks) nonlinear waves and shocks acting in times comparable to the 
observations  (see  Paper 3).

 In the bottom region of Fig.~\ref{fig:uno}, e.g.,  $y\sim 10$Mm,  we  note  
an increase  of the density (Fig.~\ref{fig:uno}a), the $\beta$--parameter (Fig.~\ref{fig:uno}d) and the total pressure (Fig.~\ref{fig:uno}e) with respect to  their corresponding  background values; correspondingly we note a  decrease of the magnetic field (Fig.~\ref{fig:uno}c). 
These bottom overdense  features  
 can be associated with the  bright
downward features described by \citet{2000SoPh..195..381M}. In our  scenario,  they  are  shock  fronts that open the way towards  the sun surface. They  modify the background medium producing, among other effects, a mass pile up  in front of the flow. A future 3D simulation could help to understand  the  relation and interaction of these reconnection outflows  with the observational bright leading loops. 

\begin{figure}%[htb!]
\centering
\includegraphics[width=7.cm]{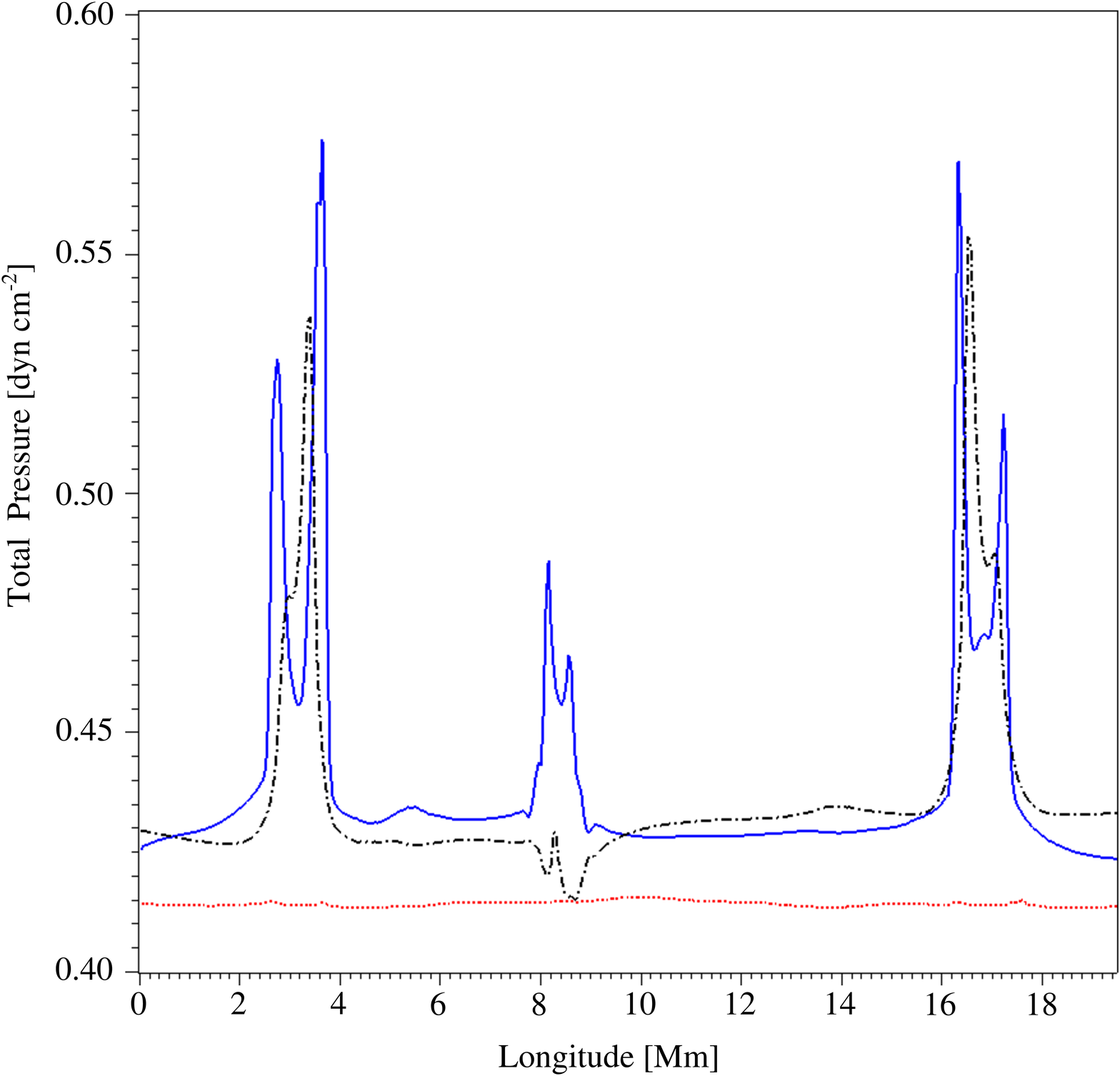}
  \caption{ Total pressure of the three outflows at $t=196$sec and heights:  $32$Mm (dotted line), $15$Mm (solid line) and $12$Mm (dashed--dotted line). See the constriction  located at $12$Mm in the second outflow trace.}
\label{fig:dos} 
\end{figure}

The multiple shock dynamic structures produce  temporal patterns which are reminiscent of observational constrictions
\citep{2003SoPh..217..267I}, \citep{2003SoPh..217..247I}, see the knots in the SARO traces of Fig.~\ref{fig:uno}. 
Figure~\ref{fig:dos} shows the total pressure at the heights indicated with cuts  in Fig.~\ref{fig:uno}e: $\sim 32$Mm, $\sim15$Mm and $\sim 12$Mm. 
Note that the total pressure is fairly constant in the upper region,  near the triggering site at $\sim 32$Mm (dotted line). At  $\sim15$Mm, (solid line) the total pressure is higher inside the tracks. The constrictions  or  knots (e.g see the second outflow trace at $\sim 12$Mm, dashed--dotted line)  correspond to regions of total internal pressure  lower than the outside values.  

As shown in Fig.~\ref{fig:uno}, in accordance with what is expected from observations, it appears  that the SARO  lanes  are not 
 distorted by the fact that they are multiple. There is not an appreciable evidence of the interaction between outflows.
  \citet{2012ApJ...747L..40S}, in the frame of  the scenario proposed by \citet{2009EP&S...61..573L}, interpreted the discreteness of outflows as an indication 
that the reconnection is highly localized.
In our scenario, as in the one proposed by \citet{2009EP&S...61..573L}, 
 the production of plasma  outflows is associated with reconnection
bursts, but we do not  assume a  2D configuration nor a particular magnetic topology  of the current sheet \citep{1995JGR...10023443P}.  
The scenario proposed by  \citet{2009EP&S...61..573L} relates inherently the plasma voids to reconnections that occur within the current sheet. 
In our case reconnection is not modeled but simulated as a pressure pulse that triggers the outflow production in an initially homogeneous media. Thus, we can still be concerned about the extent of the independent behavior between different outflows.

\subsection{Structure and wavy appearance}
An indication of the interaction between SAROs seems to be the wavy appearance that can be seen in some observations, e.g.  \citet{2005A&A...430L..65V}.
There is an important difference between 1D and 2D simulations in relation with the ability to generate the wavy appearance of the reconnection outflows.
In Papers 1-2 (1D simulations) these features are simply obtained due to lateral rebounds that emulate a neighbor denser medium (see Fig.~2 and Fig.~1 of Paper 1 and 2 respectively). The oscillatory pattern is sustained over time and the wave structure does not lose energy. In Paper 3 (2D),  the wavy appearance of a lateral cut is damped in few periods (see Fig.~1a of that paper), but if we focus in the 2D slides (see Fig.~2 of Paper 3)  it is not possible to recognize the wavy character. 
 It seems that the spatial distribution of energy of the 2D lateral rebounds (with closed boundary conditions) leads to a rapid damping of the oscillations and would not give account of the wavy appearance of the observations.

\begin{figure}%[htb!]
\centering
 \includegraphics[width=7.cm]{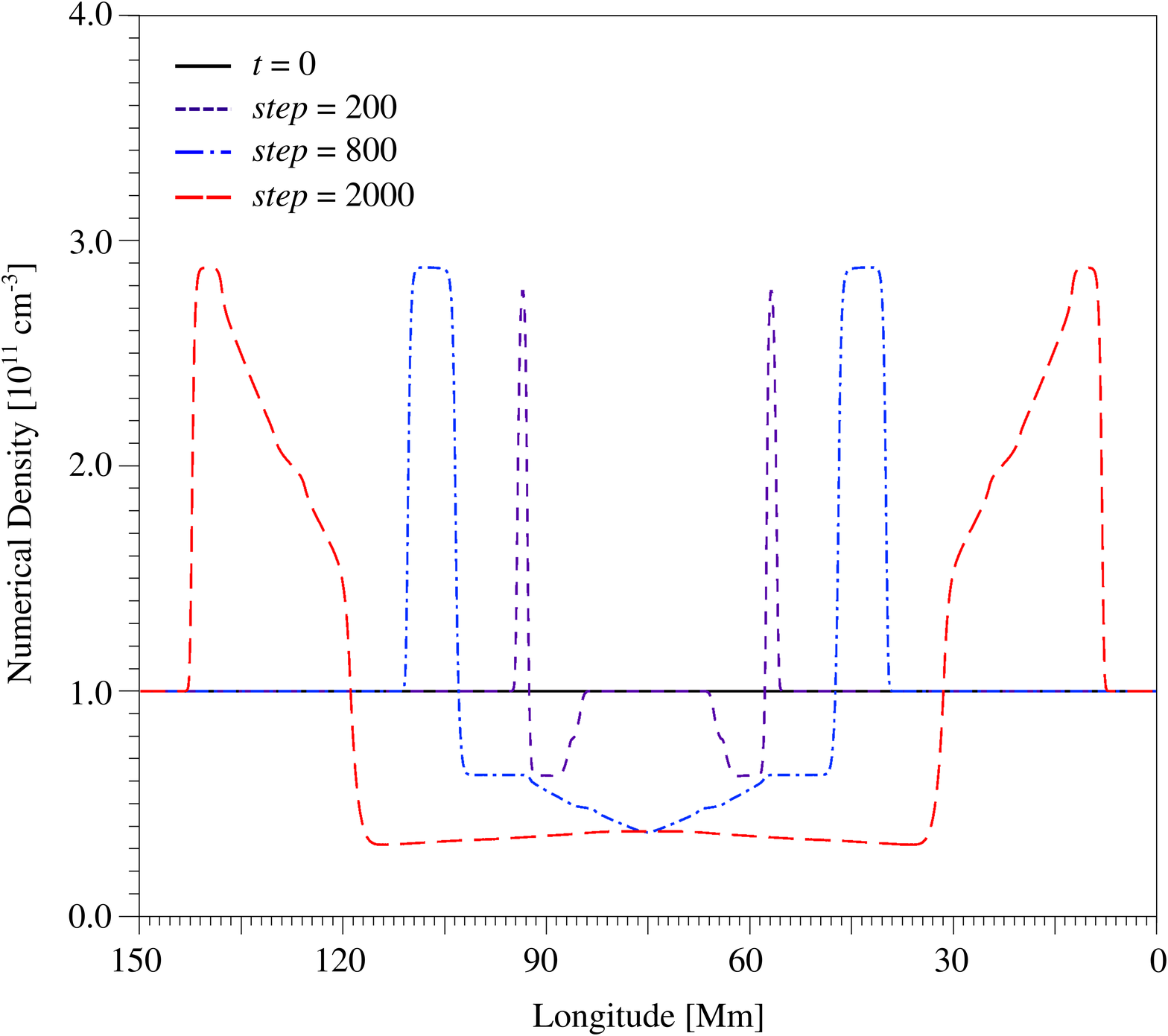}
\includegraphics[width=7.cm]{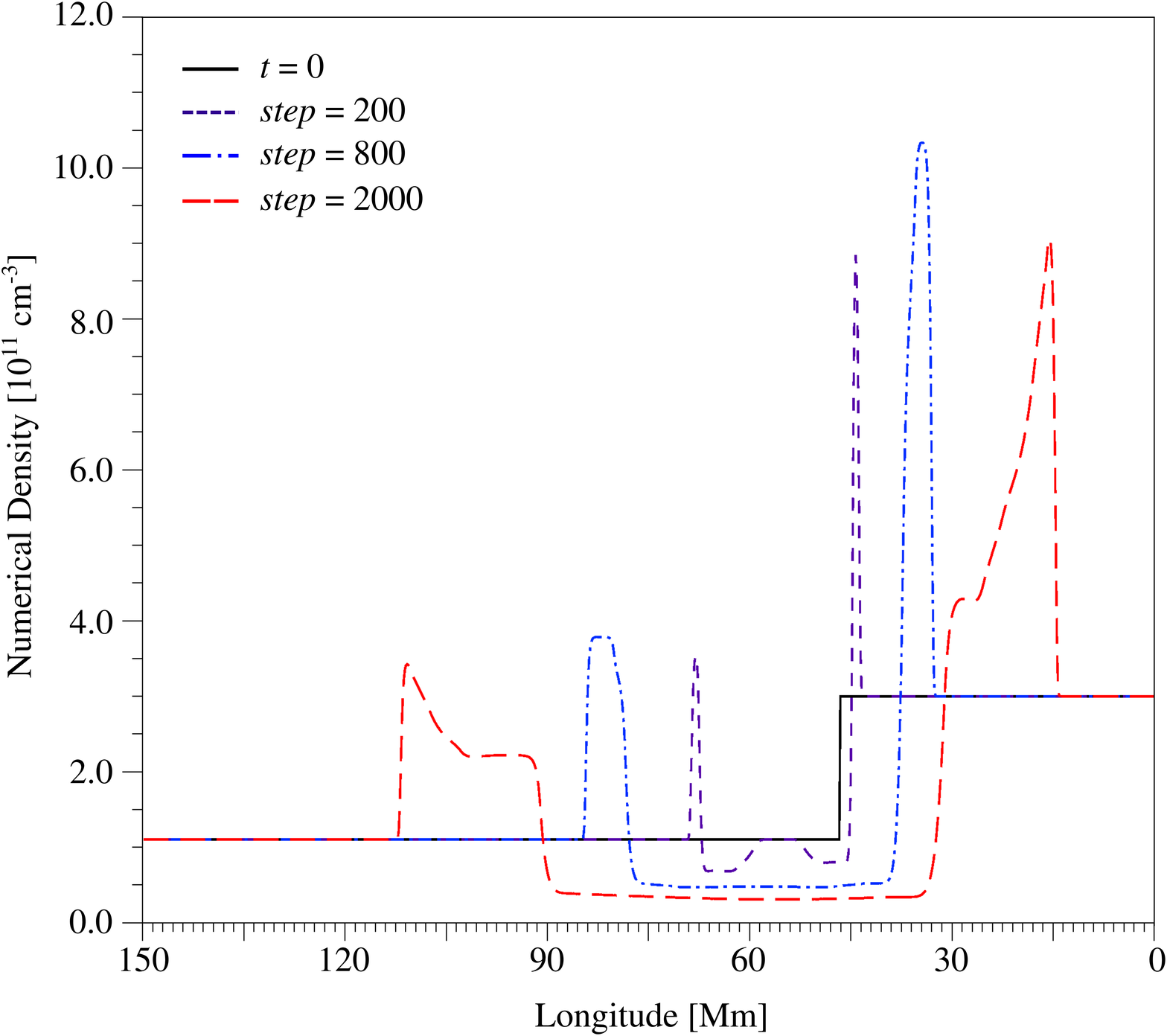}
\includegraphics[width=7.cm]{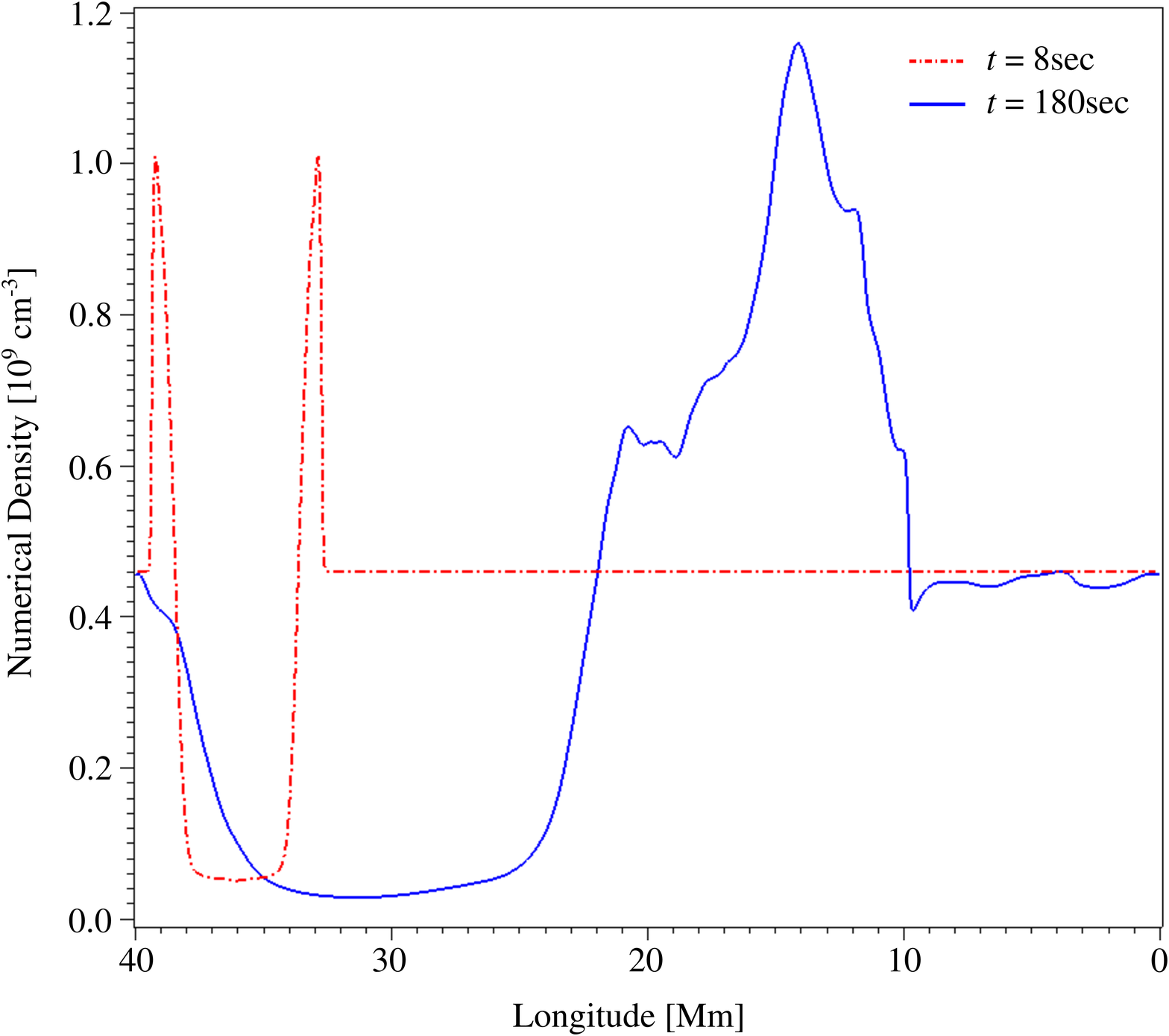}
  \caption{Upper panel) 1D density evolution after a central triggering pressure pulse for different time steps, Middle panel) 1D density evolution after an asymmetric triggering pressure pulse for different time steps, Lower panel) 2D SARO density evolution for a cut  at $x=19$Mm in Fig.~\ref{fig:uno}a.}\label{fig:tres} 
 \end{figure}

Meanwhile,   comparing the evolution of  nonlinear waves produced by a triggering pressure pulse in the 1D and 2D cases we recognize a similar dynamic behavior  that prevails during time and could  explain the robustness of the voided patterns. Figure~\ref{fig:tres}a shows the spontaneous 1D non--dimensional density evolution due to the action of a  pressure pulse over a medium with  static initial  condition  \citep{2009MNRAS.400.1821F}. 
In Figure~\ref{fig:tres}b  the triggering pulse is  asymmetrically  located, as  in our runs. 

A detailed description of the  evolution of the voided features can be obtained from these figures 
 and from the wave analysis provided by the numerical techniques used.  
Figure~\ref{fig:tres}a displays nondimensional  time step numbers $200,\  \ 800,$ and $ 2000.$   Time step $0$ corresponds to $\rho(0)=1.$ In the figure we see two  shock wave fronts for each time step, e.g., for step  $800,$ the shocks are located at    $(x_{1s}=39;  \ x_{2s}=111)$Mm. Also, for the same time step  two  contact discontinuities are found at $(x_{1c}=46.5;  \ x_{2c}=103.5)$Mm and two  expansion waves are found starting  at $(x_{1e}=57; \  x_{2e}=93)$Mm.
 Initially  two  shock fronts moving away from the triggering pressure pulse location
are formed (see e.g. time step  $200$ in Fig.~\ref{fig:tres}a and Fig.~\ref{fig:tres}b). As the shocks travel along the magnetic field direction, the temperature and the density  are increased making the energy rise, e.g. $(x_{1s}=39;  \ x_{2s}=111)$Mm in time step $800$ of Fig.~\ref{fig:tres}a.  The density is abruptly diminished (e.g. $(x_{1c}=46.5;  \ x_{2c}=103.5)$Mm in time step $800$) by  contact discontinuities that go behind the shocks and  can be recognized  because the pressure and the velocity of the flow are not changed while the waves pass. The temperature  increases to maintain a constant  pressure  across the contact discontinuities. Also,  two expansion waves --that are recognized because they diminish the density while they pass-- are initially produced and travel towards the triggering position, contrary to the shock fronts (e.g. $(x_{1e}=55.5; \  x_{2e}=94.5)$Mm at time step $200$ in Fig.~\ref{fig:tres}a).  These  waves collide, rebound at the triggering location  (see Fig.~\ref{fig:tres}a--b, center part for time steps $800$),  and then travel in the opposite direction towards the contact discontinuities (e.g. $(x_{1e}=57; \  x_{2e}=93)$Mm of time step $800$ in Fig.~\ref{fig:tres}a).   From the comparison of time step $800$  and time step $2000,$  we see that the expansion waves have a larger speed than  the contact discontinuity.
 At time step $2000$ the position of the  shock  fronts in Fig.~\ref{fig:tres}a are ($x_{1s}=7.5; x_{2s}=142.5$)Mm,
 the expansion waves have interacted with the contact discontinuities and lowered the density and pressure leaving behind a  coupled nonlinear system of waves, e.g., the ranges $(\Delta x_{1}=31.5-7.5;\Delta x_{2}=142.5-118.5)$Mm in Fig.~\ref{fig:tres}a and $(\Delta x_{1}=28.5-15.0; \Delta x_{2}=109.5-91.5)$Mm in Fig.~\ref{fig:tres}b.
As a result of this dynamic a central voided cavity is formed.

 Figure~\ref{fig:tres}c, shows a 2D density cut taken along  the SARO located at $x=19$Mm  in Fig.~\ref{fig:uno}a. The similarity between Fig.~\ref{fig:tres}a--b  and Fig.~\ref{fig:tres}c (compare $t=8$sec in Fig.~\ref{fig:tres}c with time step $800$ in Fig.~\ref{fig:tres}a--b and $t=180$sec in Fig.~\ref{fig:tres}c with time step $2000$ in Fig.~\ref{fig:tres}a--b)   expresses the likeness between the dynamic behavior of the 1D and 2D descriptions. There are, however, small differences in the shape of Fig.~\ref{fig:tres}c at $t=180$sec    from the ones in Fig.~\ref{fig:tres}a--b, time step $2000.$
 Comparing the features ranging from $x=(10-20)$Mm in Fig.~\ref{fig:tres}c  with the corresponding features in Fig.~\ref{fig:tres}a--b, we note than even when the patterns are similar the more irregular shape in 
  Fig.~\ref{fig:tres}c can be attributed to perturbations produced by the other  SAROs.

\begin{figure}[htbp!]
\centering
 \includegraphics[width=7.cm]{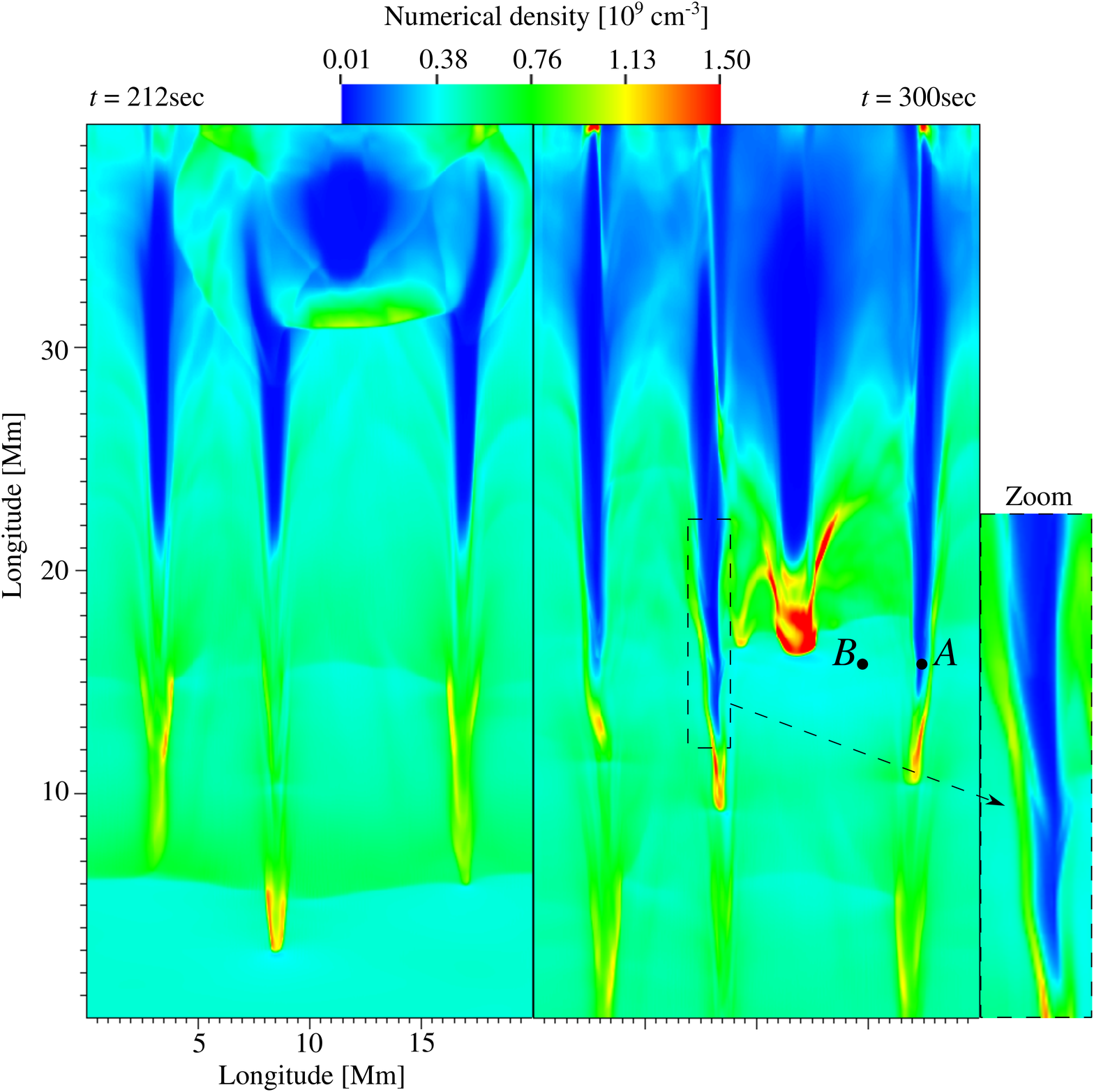}
 \caption{ $M1$ numerical density after a fourth pressure pulse, the event is seen at left) $t=212$sec, and at right) $t=300$sec. Note the wavy features (zoom) due to the interaction between SAROs. }
\label{fig:cuatro} 
\hfill
\includegraphics[width=7.cm]{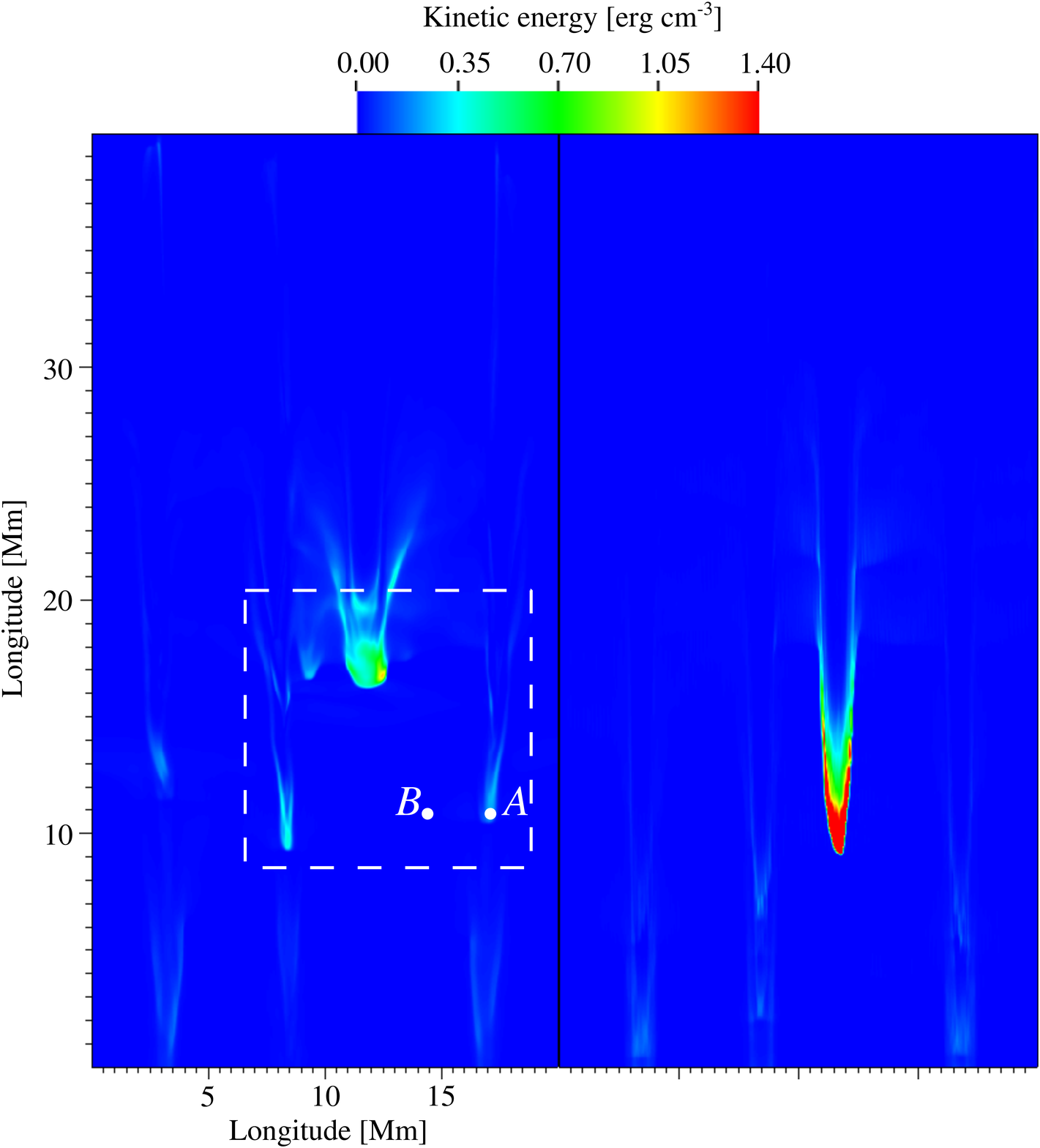}
 \caption{ Kinetic density energy collimated into the wakes left by previous SAROs; left) $M1$ at  $t=300$sec, and right)  $M2$ at $t=300$sec. }
\label{fig:cinco}
\end{figure}

However, we found cases  where the interaction between  SAROs is evident and has the observational consequence that it reproduces the wavy  character registered by \citet{2005A&A...430L..65V}. Figure~\ref{fig:cuatro}a-b
shows the SARO evolution  for $M1$ after a new  pressure pulse,   triggered at $t=200$sec, has developed a wavefront over
  the discrete tracks left by the three initial SAROs that have
traveled decelerating a distance of $\sim 16$Mm
 (Fig.~\ref{fig:cuatro}a).
 In
 the upper part of Fig.~\ref{fig:cuatro}b at  $t=300$sec it appears that the initial SAROs have lost
 their discreteness. We have stated that when the magnetic field is uniform the  energy produced by the pressure pulse is distributed in a non--homogeneous way due to the collimation generated by the magnetic field. While the SAROs propagate further  a new trace left by the new burst 
 can be seen 
 sunwards  (Fig.~\ref{fig:cuatro}b). Also, the new reconnection pulse injects energy most of  which is distributed in a non--uniform  way between the wakes left by the previous SAROs i.e.,
the initially spherical pressure pulse (Fig.~\ref{fig:cuatro}a) at $212$sec is later  collimated by the magnetic field configuration at $300$sec (Fig.~\ref{fig:cuatro}b). As in the other cases, the sunward elongation of the fourth lane  is due  to the interaction of upper rebounds 
and downward absorptions along the magnetic field that collimates the flow.  

A remarkable effect which has observational implications is the 
distribution of kinetic energy. As can be seen from Fig.~\ref{fig:cinco}a, part of the new SARO kinetic energy is  distributed along the wakes left 
by the previous SAROs. In accordance with the observations, when the sequence is seen as a movie  the wakes   exhibit a wavy  motion  
(e.g. \citep{2005A&A...430L..65V},  \citep{2000SoPh..195..381M}, \citep{2007A&A...475..333K}). This can be  explained due to the tendency of the 
wavefront energy to deviate into lower density regions, i.e., the wakes left by the previous SAROs. 
The continuity Rankine--Hugoniot relation for 
hydrodynamic shocks, i.e., the shock in the longitudinal direction (Paper 1--3), and in the  shock wave frame is
$$v_{u}=\frac{\rho_{d}}{\rho_{u}}v_{d} \ \  \ \  \ \ u \ \  \ \ upstream,  \ \ d  \ \ downstream.   $$
Initially, the energy of the pressure pulse is almost spherically distributed due   to its larger pressure value with respect to the ambient one (Fig.~\ref{fig:cuatro}a), so we can consider that $\rho_{d}$ and $v_{d}$ are almost the same all around the spherical  boundary of the initial shock front. However, the corresponding upstream $\rho_{u}$ values will be different depending on the previous SARO distribution, i.e., $\rho_{u}$ is almost an order of magnitude lower in the wakes left by the previous SAROs. Thus, initially the upstream flow speed will be larger in the SARO wakes than  in the neighbor ambient. 
Moreover, the previous SAROs have larger values of the flow speed than the ambient one and its value will be added to the upstream speed.
This implies that two upstream  shock front positions with initial values $\rho_{A}<\rho_{B},$ and thus  $v_{A}>v_{B},$ will continue their motion with  $v_{A}>v_{B}$ while the shock wave travels. Obviously, the same relation  stands for the  kinetic 
energy, i.e.  $K_{A}=\rho_{A} v_{A}^{2}=\rho_{B} v_{B}v_{A} >\rho_{B} v_{B}^{2}= K_{B}$ ($K$ the kinetic energy). Upstream, this relation is preserved as can be appreciated in Fig.~\ref{fig:cinco}a. Compare the kinetic energy of $A$ and $B$ in this figure with their  corresponding densities  ($\rho_{A}<\rho_{B}$) in Fig.~\ref{fig:cuatro}b. 
 Note the correspondence between the low--high density values in  positions  $A$ and $B$  in Fig.~\ref{fig:cinco}a with respect to the high--low kinetic energy values in the same positions in  Fig.~\ref{fig:cuatro}b. The energy of the fourth reconnection event is 
partly  deviated from the new wake  into the previous ones depending on the background plasma parameter. Figure~\ref{fig:cinco}b
shows the kinetic density energy for $M2$ at $t=300$sec. 
When the magnetic field intensity is increased (or the intensity of the pressure pulse is diminished, model $M4$) the wavy character is lost. 

\subsection{A possible SAD configuration}

\begin{figure}[ht!]
\centering
 \includegraphics[width=6.cm]{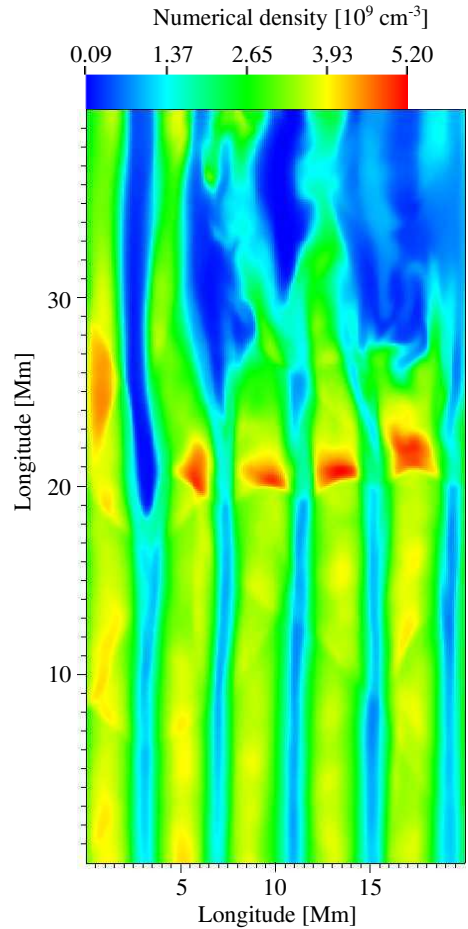}
  \caption{New reconnection events occurring in a medium distorted by previous SAROs.}\label{fig:seis} 
 \end{figure}

We performed a new simulation with a non--homogeneous  background that emulates the modification of the medium due to the action of several evolved SAROs. The previously simulated   thin  dark tracks (of sizes  $ \leq 2$Mm),  emulate the action of  reconnection pulses in a homogeneous medium.  
The bright overdense shockwave fronts seen at the bottom  of Fig.~\ref{fig:uno} would lead to a mass pile up in front of the flow changing the background density. To analyze the effect of  new reconnections triggered in a medium modified by the SAROs, and that resembles the fan ray features described by the observations, we use a sinusoidal initial background density conditions. Figure~\ref{fig:seis}
shows the  density variations    obtained from the evolution of four  pressure pulses located randomly in the medium  distorted by previous SAROs. 
As shown in  Fig.~\ref{fig:cinco} the energy of new reconnection events is confined preferentially into the lanes left by previous reconnection events,  i.e., thus the new dark voids are seen trailing brighter  lanes (SAROs) once the medium is non--homogeneous. From  Fig.~\ref{fig:seis} we note that  the later pressure pulses, of a teardrop shape, are associated with different sizes ($\sim  [2-9]$Mm)  depending on the effect produced 
by the new wavefront in  the medium.  Note the density range variation and the occurrence of bright and dark lanes that resemble  the features seen in Movie 2 provided by \citet{2012ApJ...747L..40S} (see e.g. the 131 bandpass slice at $12:08:33$UT). Also, the bottom slice at  $12:22:09$UT (Movie 2) exhibits a complex structure that resembles the action of evolved SAROs,  where some loops are also recognized at the left bottom. If this scenario describes a possible plasma configuration, SADs could be late reconnection events that occur in a non--homogeneous medium distorted by the action of the previous localized SAROs and, the linking between  SAROs and SADs would be related with the fact that energy of later reconnections is confined into the lanes left by the formers. 
This new burst of kinetic energy, suddenly confined into the older lanes, renews the accelerated motion of the SAROs and deforms their extremes until they acquire a teardrop--shape with an oscillating tail. Thus, these SADs would be seen trailing SAROs. This scenario  does not exclude the possible  presence and interaction between loops and SAROs as a result of the magnetic field reorganization, however, this needs of a 3D simulation that is beyond the scopes of this paper.  An important difference with  the scenario proposed by \citet{2012ApJ...747L..40S} is that all the obtained simulated voids, some of them  resembling SADs, are not wakes but  reconnection outflows formed by shocks and nonlinear interactions that could also be preceded by loops.  In summary, our interpretation implies that  SAROs and SADs are both reconnection outflows of the same type, excepting  that  SAROs  are previous, and they modify the medium. Later reconnection outflows will give  the appearance that they are preceded by brighter lanes. What is lacking is to understand the interaction between SAROs and probable nearby loops.

\section{Conclusions}
We analyze the dynamic of multiple dark supra--arcade lanes that move decelerating towards the sun surface. 
In the scenario we  proposed,   dark lanes are explained as reconnection outflows (SAROs) resulting from the evolution of shocks 
and expansion waves that form confined voided cavities --of high temperature and $\beta$ values-- collimated in the direction of the ambient magnetic field. These features sustain their morphology over time behaving as a whole pattern (almost independently from boundary conditions and other perturbations) with a sunward and decelerating motion triggered by upward reconnection events.

When multiple SARO configurations are analyzed  we note that the structure of each SARO is similar to the individual pattern reinforcing   the non--interacting hypothesis. However, we found that the wavy character that can be seen in  observations can be interpreted as  an indication of interaction between SAROs. This interaction is significant when the bursts that trigger the phenomenon act over the lanes left by previous SAROs. This wavy character is enhanced  with the strength of the reconnection and/or with lower values of the magnetic field intensity.

The fact that in a non--homogeneous medium i.e., modified by reconnection outflows (SAROs), the energy of new reconnection events is collimated into the voided lanes left by the previous SAROs implies  a linking between the previous and later  reconnection events that can be consistent with the observations. 
\section{Acknowledgments}
We are thankful to an unknown referee who helped us to improve the paper. These paper was supported by the CONICET grants PIP N° 112-200801-00754 and PIP N° 200801-02773.
 \\ 
\bibliography{MScecereR2}
\end{document}